# Tunneling into the normal state of $Pr_{2-x}Ce_xCuO_4$

Y. Dagan, M. M. Qazilbash and R. L. Greene.

*Center for Superconductivity Research, Department of Physics, University of Maryland, College Park, Maryland 20742, USA.*

The temperature dependence of the tunneling conductance was measured for various doping levels of $Pr_{2-x}Ce_xCuO_4$ using planar junctions. A normal state gap is seen at all doping levels studied, x=0.11 to x=0.19. We find it to vanish above a certain temperature T*. T* is greater than $T_c$ for the underdoped region and it follows $T_c$ on the overdoped side. This behavior suggests finite pairing amplitude above $T_c$ on the underdoped side.

In conventional superconductors, below a certain temperature $T_c$, pairs of opposite spins and momenta are formed. At that temperature, these pairs also condense into a state of zero resistivity when an energy gap, $\Delta$, opens up in the density of states. This seems not to be the case in the high-$T_c$ cuprates. For the hole doped cuprates Renner *et al.*[1] have shown that a gap in the density of states persists above $T_c$ especially for underdoped cuprates. This behavior was referred to as the pseudogap state.[2] One of the interpretations put forth to explain this phenomenon was in terms of incipient superconductivity above $T_c$ with a lack of phase coherence.[3] Another possibility is that the pseudogap is due to some order not necessarily following $T_c$ on the overdoped side.[4]

On the electron-doped side of the phase diagram tunneling experiments showed a gap in the density of states when the normal state was accessed by applying magnetic fields higher than the minimal field needed to quench superconductivity, $H_{c2}$ [5, 6]. Alff *et al.*[7] later showed that this gap vanishes at a doping dependent temperature, (T*) lower than $T_c$ for doping levels close to the optimum. This appeared to rule out the possibility of preformed pairs in the electron-doped cuprates and suggested that this gap was related to an order parameter that competes with the superconducting one. Recently, we found evidence for a quantum phase transition near x=0.165±0.005 from transport measurements on $Pr_{2-x}Ce_xCuO_4$ (PCCO).[8] It is tempting to relate this phase transition with the gapped spectra observed in tunneling. However, unlike the hole doped cuprates, in the electron-doped cuprates a broad region of antiferromagnetism in the phase diagram extends up to very high dopings, possibly into the superconducting dome. [9,10] This suggests that the quantum phase transition we have observed is likely to be an antiferromagnetic (AFM) to paramagnetic transition. On the other hand, we are going to show here that the normal state tunneling gap (NSTG) persists even in the highly overdoped x=0.19 sample, i.e. way beyond the AFM region. This suggests that the NSTG is unrelated to the Quantum phase transition we previously reported and probably not directly to the AFM phase.

In this letter we report an extensive and systematic tunneling study into PCCO as a function of field and temperature for many Ce doping levels: x=0.11, 0.13, 0.15, 0.16, 0.17, 0.18 and x=0.19. We find the NSTG temperature, $T^*$, to be greater than $T_c$ for x<0.17 and T*≈$T_c$ for x≥0.17. This is a very different behavior than found previously [7]. Our data suggests that pairs are formed and condensed at the same temperature for x≥0.17, as in conventional superconductors.

Lead (Pb) contacts, approximately 5000Å thick were evaporated on fresh PCCO films and on faces parallel to the *c*-axis (*ab* faces) of freshly cleaved $Pr_{1.85}Ce_{0.15}CuO_4$ single crystals. The junctions' area was approximately $0.5\times0.5mm^2$ for the films and $1mm\times30\mu m$ for the crystals. This results in a true tunneling contact. In both cases, the PCCO surface is exposed to ambient atmosphere for not more than a few minutes. It has been demonstrated in $Al/YBa_2Cu_3O_{7-\delta}$ contacts that the aluminum counterelectrode takes oxygen from the cuprate thus creating an oxide barrier [11]. A similar process occurs in PCCO/Pb contacts [12]. To prevent over-reduction of oxygen in the vicinity of the junction we use slightly oxygen rich films. In such a case there could be a slight difference of oxygen concentration between the bulk film and the junctions' area. For that reason $T_c$ report here is measured at the junction using the tunneling characteristics itself. The bulk $T_c$ measured in resistivity and the $T_c$ measured by the tunneling conductance differ by less than 2K. I(V) characteristics were measured using a current source and a voltmeter and were differentiated digitally. The conductance characteristics of junctions made on films and on crystals of similar doping level show no difference. This indicates that tunneling into the films, although nominally *c*-axis oriented, is predominantly in-plane tunneling due to *ab* plane facets exposed to Pb. The magnetic field is always applied along the *c*-direction.

In Fig.1 we show the conductance versus voltage for three different doping levels at 1.8K. The black solid lines are zero field measurements. We note the low conductance at zero bias, indicates good tunneling contacts with negligible leakage currents. The strong Pb and PCCO coherence peaks and the Pb phonons can be clearly seen. The red lines are measurements taken above the Pb critical field (0.1-0.12T). Only the PCCO coherence peaks can be seen and the Pb phonons are absent. At $H>H_{c2}$ (blue line) superconductivity in the PCCO electrode is quenched; yet a small normal state gap (NSTG) feature remains. It is seen as a depression of the conductance at zero bias. It exists in *all* doping levels studied.

Since the x=0.11 sample is not superconducting we are able to follow the magnetic field dependence of the NSTG from 0.12T (when the Pb becomes normal) to 14T as shown in Fig. 2a. In fig.2b we show the conductance at zero bias as a function of magnetic field applied parallel to the *c* direction. The NSTG partially closes at low fields, saturating above ~8T. The low field range where most of the change occurs is hidden by the superconductivity in the other samples. We are not aware of theoretical prediction for such a field-dependence. Further experimental and theoretical studies are needed in order to find out the origin of this field dependence.

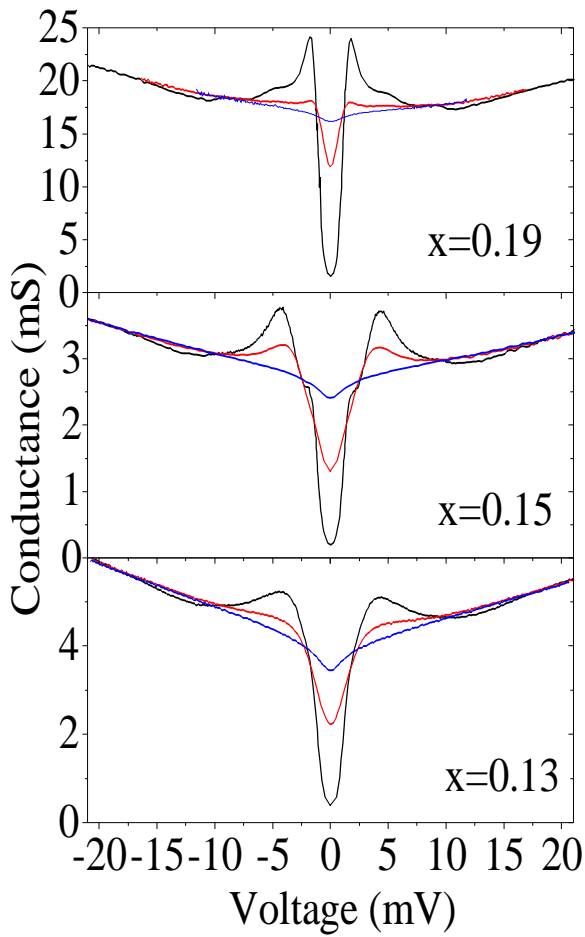

**Figure 1.** The conductance *versus* voltage for Pb/PCCO junctions for three typical dopings: x=0.13, x=0.15, x=0.19 at T=1.8K. The magnetic field is applied parallel to the *c*-axis: Black line H=0, red line $\mu_0H=0.12T$ (larger than the Pb critical field), blue line $\mu_0H=14T$ (both electrodes are normal). Note the normal state gap in all three doping levels.

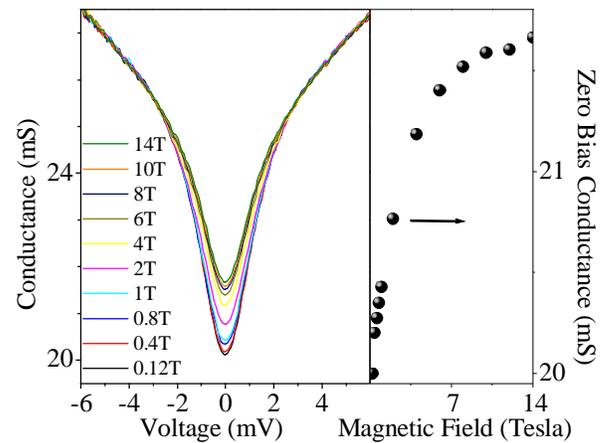

**Figure 2.** Nonsuperconducting $Pr_{1.89}Ce_{0.11}CuO_4$ sample T=1.8K (a) conductance versus voltage at various fields (b) Zero bias conductance vs. magnetic field taken from Figure 3(a). The normal state gap exhibits magnetic field dependence saturating above 8T.

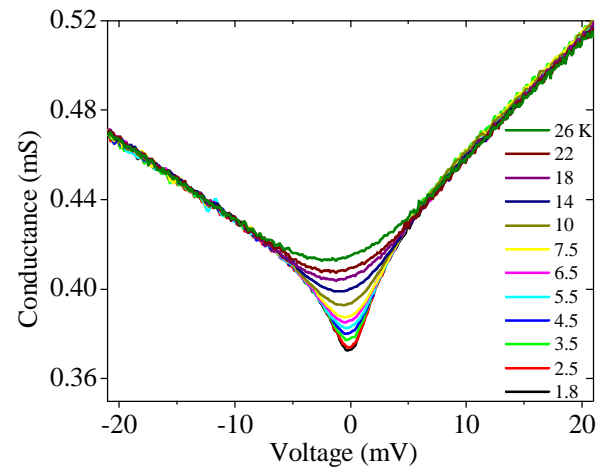

**Figure 3.** The conductance versus voltage for $Pr_{1.85}Ce_{0.15}CuO_4$ crystal at 12T along *c*-axis and various temperatures.

The temperature dependence of the NSTG at 12T for x=0.15 single crystal is shown in fig. 3. It can be seen that the NSTG closes with temperature. A linear background independent of temperature is observed at high biases. We used this background and its extrapolation to zero bias for the analysis of the temperature dependence of the NSTG as described below. We made conductance *versus* voltage measurements at various temperatures for all doping levels studied. The NSTG temperature T* is determined by analyzing the temperature dependence as depicted in fig.4. The temperature has two effects on the conductance: The first effect is thermal smearing resulting from the broadening of the Fermi function with increasing temperature. The second effect is a possible decrease in the NSTG amplitude. Both effects result in decreasing the gap features. To isolate the latter effect we do the following: at any temperature (10K is shown as an example) we calculate, taking into account thermal broadening effects, the expected conductance at (say) 10K assuming the NSTG is constant with temperature. In this case the temperature dependence of the conductance can be calculated using

$$g(eV,T_1) \propto \int g(E,T=0)\frac{\partial f(T_1,E-eV)}{\partial E}dE$$ (dashed black line in Fig.4a). Since we do not have the zero temperature conductance data we use the 1.75 K data instead. This is a slight overestimation of the thermal smearing since the 1.75K data is already somewhat smeared. As a background we use the linear approximation (see Fig.3), which is the simplest possibility for such a background. Plugging this linear background (at T=0) into the thermal smearing integral above, we obtain the background at any temperature needed (the temperature affects only the low bias region). For example at 10K for the x=0.15 sample one obtains the green line in fig. 4a. Now, for example, at 10K (Fig. 4a) we compare the real measurement (dotted red line) with the 1.75K data smeared to 10K (dashed black) and the linear background smeared to 10K (solid green). We define the relative depth of the NSTG as the ratio a/b (see Fig. 4a) and we follow it as a function of temperature. It decreases monotonically to zero (See figure 4b) with increasing temperatures. The temperature at which the ratio a/b goes to zero is defined as T*. In Fig 4b we show the temperature dependence of the relative depth of the NSTG for two doping levels: x=0.15 and x=0.19. We did a similar analysis for all the doping levels under study. The results for T* are summarized in Fig.5 (black circles).

We also show in Fig.5 (red squares) the critical temperature, $T_c$. A slight oxygen variation between the bulk film and the vicinity of the junction may result in a small difference between the bulk transition temperature and the local $T_c$ at the junction. This variation is due to oxygen taken from the PCCO film to create the oxide barrier. Therefore, the superconducting transition temperature, $T_c$, was measured at the junction by comparing, at each temperature, the zero field tunneling measurement to the 14T spectrum. $T_c$ was defined as the temperature at which these two spectra overlap. We note that at low doping levels T*>$T_c$, while these two temperature scales become closer around optimum doping and follow each other on the overdoped side. This behavior is totally different than that reported by Alff *et al.*. [7] Their T* was found to be lower than $T_c$ in the vicinity of optimum doping extrapolating to zero at x=0.17.

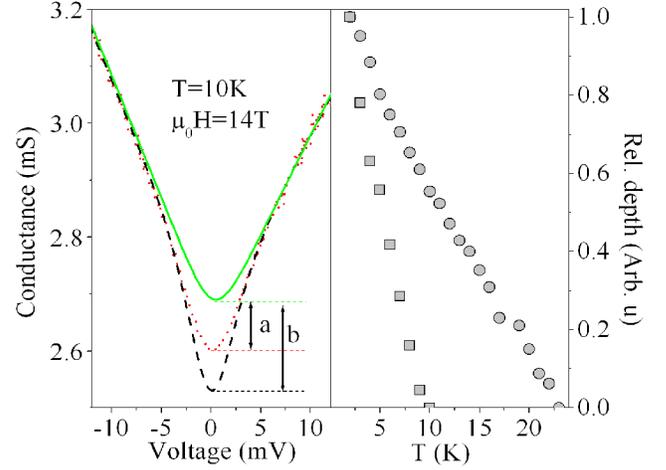

**Figure 4**. (a) $Pr_{1.85}Ce_{0.15}CuO_4$ junction conductance. Black dotted line is the 1.8K data smeared to 10K assuming that thermal smearing is the only effect of temperature. Red dotted line is the real data at 10K. Green solid line: linear background extrapolated from the high bias measurement and thermally smeared to 10K. The ratio a/b is defined as the relative depth of the gap (b) The relative gap depth as a function of temperature for x=0.15 sample (circles) and for x-0.19 samples (squares) T* is defined as the temperature at which the relative depth goes to zero.

We shall now discuss the possible origin of the NSTG. Biswas *et al.*[6] suggested that the NSTG results from electron-electron interactions. This should manifest itself suppression of the tunneling density of states at zero bias, with a conductance which is logarithmic in voltage at low biases. [13] However, if electron correlations are responsible for the NSTG it is rather surprising that T* follows $T_c$ on the overdoped side. Moreover, such correlations also result in a ln(T) conductivity [14]. By contrast, in our overdoped films a metallic like conductivity is observed [8] where the NSTG is still seen.

A second possibility is that the NSTG is due to a partial gapping of the Fermi surface in the AFM normal state characterizing PCCO [9,15]. Tunneling into the AFM state of Chromium showed conductance with weak features at zero bias [16] or sometimes even featureless conductance [17]. If the NSTG we observe is due to antiferromagnetism it is puzzling why we still observe it at doping levels higher than x=0.17 where neutron scattering experiments show no antiferromagnetism. [10]

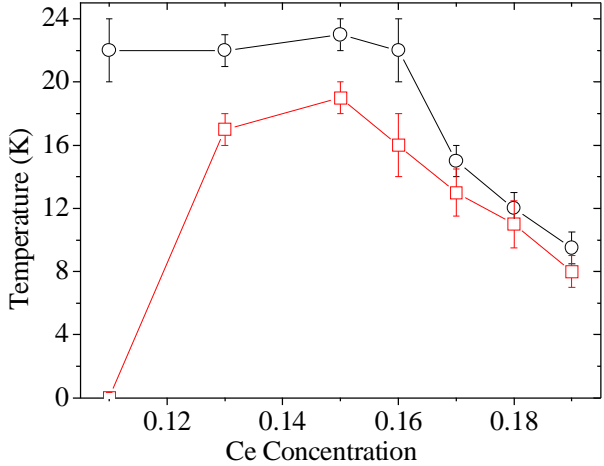

**Figure 5.** Black circles: T*, the temperature at which the normal state tunneling gap vanishes, determined using the analysis described in the text. Red squares: $T_c$ of the junction (see text for detail).

A third possible explanation for the NSTG is nonequilibrium tunneling due to nonzero electron relaxation times [18]. However, the depression of the zero bias conductance in that case is typically of the order of 0.3%, much smaller than the feature we report here.

Alff *et al.*[7] suggested that the pseudogap in the electron doped cuprates is due to a hidden order under the superconducting dome. Their data suggested that T* goes to zero around x=0.17. Recently, we showed evidence for quantum phase transition in PCCO at a critical doping x=0.165±0.005.[8] The data presented by Alff *et al.* appear to be consistent with such a transition, However, the extensive doping study and the simpler data analysis we present here suggest that the NSTG is not a signature of the order parameter that vanishes at $x_c$=0.165. This vanishing order is more likely to be antiferromagnetism as suggested by other experiments. [9, 10, 15].

Our data shows T*≥$T_c$ in contrast with Alff *et al.*[7]. The data we present here suggests a pre-formed singlets scenario. A non vanishing pairing amplitude exists up to $T^* > T_c$ in the underdoped region of the phase diagram while $T^* \approx T_c$ in the overdoped region. This is consistent with previous Nernst effect data, suggesting broader fluctuation region on the underdoped side [[19]]. Unlike the hole doped cuprates, T* does not increase when the doping is decreased from the optimum level. [20] This can be explained by the proximity of the AFM phase possibly persisting into the superconducting dome. The AFM channel competes with the pairing channel and hence the saturation in $T^*$.[21] However, we note that although the NSTG vanishes at T*≈$T_c$ on the overdoped side its field dependence is rather different from the superconducting gap. While the superconducting gap vanishes at H>$H_{c2}$ the NSTG still persists up to fields as high as 14T. This needs to be understood in the future.

In summary, we measured in-plane tunneling conductance into $Pr_{2-x}Ce_xCuO_4$ films and crystals. A normal state tunneling gap (NSTG) is seen in the whole doping range studied x=0.11-x=0.19. We studied its doping, temperature and field dependences. In the underdoped region the NSTG exists above the superconducting transition temperature, $T_c$. T*, the temperature at which the NSTG appears, merges into $T_c$ on the overdoped side. Finite pairing amplitude above $T_c$ on the underdoped side is the most plausible explanation for our data.

We thank A. V. Chubukov, A.J. Millis and G. Deutscher for discussions; R. Beck for computer simulations; NSF grant DMR-0352735 supported this work.